\documentclass[showpacs,showkeys,preprintnumbers,pre,amsmath,amssymb,superscriptaddress,twocolumn]{revtex4}
\usepackage[english]{babel}
\usepackage[utf8]{inputenc}
\usepackage[colorinlistoftodos, color=green!40, prependcaption]{todonotes}
\usepackage{color}
\usepackage{dcolumn}
\usepackage{soul}
\usepackage{tabularx}

\usepackage[colorlinks=true,linkcolor=blue,urlcolor=blue,citecolor=blue]{hyperref}
\usepackage{bm}

\begin{document}

\title{A note on the lattice momentum balance in the lattice Boltzmann interaction-framework}

\author{Francesca Pelusi}
    % \email{francesca.pelusi@cnr.it}
    \affiliation{Istituto per le Applicazioni del Calcolo, CNR, Via Pietro Castellino 111, 80131 Naples, Italy}
    
\author{Matteo Lulli}
    \email{mlulli@phy.cuhk.edu.hk}% Your name
    \affiliation{Department of Physics, The Chinese University of Hong Kong, Sha Tin, Hong Kong, China}

\author{Christophe Coreixas}
    %\email{coreixas@sustech.edu.cn}% Your name
    \affiliation{Institute of Advanced Study, BNU-HKBU United International College, Zhuhai, Guangdong 519088, China}
    \affiliation{Department of Computer Science, University of Geneva, 1204 Geneva, Switzerland}

\author{Mauro Sbragaglia}
    %\email{sbragaglia@roma2.infn.it}
    \affiliation{Department of Physics \& INFN, Tor Vergata University of Rome, Via della Ricerca Scientifica 1, 00133 Rome, Italy.}

\author{Xiaowen Shan}
    %\email{xiaowenshan@uic.edu.cn}
    \affiliation{Institute of Advanced Study, BNU-HKBU United International College, Zhuhai, Guangdong 519088, China}

\date{\today} % Leave empty to omit a date

\begin{abstract}
In this note, we show how the exploitation of the \emph{lattice momentum balance} condition allows to envisage an analytical procedure to define the \emph{lattice pressure tensor} (LPT) for the multi-phase Shan-Chen (SC) lattice Boltzmann method (LBM) with single-range potential. This construction ensures that the LPT normal component to a flat interface is constant to machine precision on each lattice node, i.e., it exactly implements the mechanical equilibrium condition on the lattice. We demonstrate the robustness of the approach by providing analytical expressions for the coexistence curves for different choices of the pseudo-potential and forcing schemes in the SC-LBM. This paper offers a novel, rigorous perspective for controlling the LPT in the SC-LBM, paving the way for its application in more general settings.
\end{abstract}
\maketitle
%%%%%%%%%%%%%%%%%%%%%%%%%%%%%%%%%%%%%%%
\section{Introduction}\label{sec:intro}
%%%%%%%%%%%%%%%%%%%%%%%%%%%%%%%%%%%%%%%
Multi-phase flows are of paramount importance in various disciplines, involving both fundamental and applied aspects~\cite{brennen2005,kolev2005multiphase,balachandar2010turbulent,crowe2011multiphase,yadigaroglu2017introduction}. The associated problems feature the motion of interfaces separating bulk phases, evolving and deforming due to the action of the flow fields. The resulting non-linear dynamics has posed various challenges for the theoretical and numerical modeling, encompassing several approaches ranging across scales~\cite{chen2006computational,yeoh2019computational,prosperetti2009computational,tryggvason2011,succi2018lattice}. Among the latter, the lattice Boltzmann method (LBM) tackles the problem from a mesoscale point of view, by integrating the dynamics of discrete particle distribution functions on a lattice, whose coarse-grained description results in multi-phase hydrodynamical behavior with diffused interfaces~\cite{succi2018lattice,kruger2017lattice}. 

Starting from the first LBM applications for multi-phase flows~\cite{Gunstensen91,Grunau93,ShanChen93,ShanChen94,Swift95,Swift96}, the method experienced rapid growth with a plethora of developed applications~\cite{Huang15,kruger2017lattice,succi2018lattice}. This paper focuses on the Shan-Chen (SC) model in the context of LBM for multi-phase flows~\cite{ShanChen93,ShanChen94,ShanDoolen95,ShanDoolen96}. The SC method considers LBM dynamics equipped with lattice interactions evaluated on the lattice nodes via specific pseudo-potentials, which are local functions of the density field~\cite{ShanChen93,ShanChen94}. Multiple applications of the SC-LBM have been envisaged over the years~\cite{Shan06,YuanSchaefer06,Sbragaglia07,Falcucci07,HyvaluomaHarting08,ZhangTian08,falcucci2008lattice,Shan08,Benzi2009,Falcucci10,SbragagliaShan10,Huang11,jansen2011bijels,colosqui2012mesoscopic,Frijters12,Sbragaglia2012,Benzi2013,SegaSbragaglia13,Chen14,Belardinelli15,dollet2015two,Khajepor2015,Liu16,Xue18,lulli2018metastability,milan2018lattice,chiappini2019,Frometal19,pelusi2021,lulli2022mesoscale,pelusi2024a,pelusi2024b,yu2025filter}. An important aspect of the SC methodology is represented by the computation of the pressure tensor, whose knowledge is crucial in assessing the properties of the non-ideal interfaces~\cite{RowlinsonWidom82}. The SC-LBM is based on lattice forces; hence, the pressure tensor must be constructed once they are assigned. 

After the pioneering analysis in the seminal paper by Shan \& Chen~\cite{ShanChen94}, various studies have been made to compute the pressure tensor for the SC-LBM~\cite{Sbragaglia07,Shan08,SbragagliaShan10,LI_PRE_86_2012,SbragagliaBelardinelli13,LYCETT_BROWN_PRE_91_2015,Li_2016,Frometal19,From2020}. Some of these works define a pressure tensor through a Taylor/Chapman-Enskog expansion~\cite{Sbragaglia07,LI_PRE_86_2012,LYCETT_BROWN_PRE_91_2015,Li_2016}; however, the latter %asymptotic analysis is not able to 
studies only partially quantify deviations induced by the numerical discretization.
Moreover, following a Taylor/Chapman-Enskog expansion approach may introduce ambiguities since any continuous pressure tensor is only defined up to a symmetric tensor with vanishing divergence. Given the discrete nature of the LBM, having the pressure tensor as a lattice quantity is paramount. Indeed, other studies tackled the problem of computing {\it lattice pressure tensor} (LPT) for the SC-LBM by applying mechanical balance arguments applied directly on the lattice, without invoking any multi-scale expansion~\cite{Shan08,SbragagliaShan10,SbragagliaBelardinelli13,Frometal19,Lulli_2021}. Nonetheless, the computation of the LPT, as presented so far in the literature~\cite{Shan08,SbragagliaBelardinelli13,Frometal19,Lulli_2021} is somewhat convoluted, involving geometrical constructions that might not be straight-forward in the three-dimensional case. Moreover, the geometrical constructions invoked lack some clear connection with the LBM evolution dynamics, leaving the LPT's rigorous derivation from the LBM quite elusive. 

In a recent work, Guo~\cite{Guo_2021} analyzed the {\it lattice momentum balance} from the LBM, identifying force imbalance contributions giving rise to discretization errors. As a result, well-balanced LBM schemes were developed, offering improved descriptions of interface properties, particularly by reducing numerical errors to machine-precision fluctuations for non-moving interfaces.
Interestingly, the lattice {\it momentum balance} proposed by Guo can be seen as a restriction of the \emph{equivalent equations} framework to the momentum equation. More precisely, the \emph{equivalent equations} are macroscopic equations at the lattice level, derived from a Taylor expansion of the lattice Boltzmann equation under moment constraints similar to those in the Chapman-Enskog expansion~\cite{DUBOIS_CAMWA_55_2008,DELLAR_ICMMES_2022,BELLOTTI_NM_152_2022}. This derivation facilitates the identification of numerical errors observed for link-wise boundary conditions~\cite{DHUMIERE_CMA_58_2009} and spatially varying forces~\cite{SILVA_CF_203_2020}, or caused by hyperviscosity effects related to the collision model~\cite{wissocq2022hydrodynamic}.

In this paper, we show how the characterization of the momentum balance condition in the SC-LBM allows us to identify a rigorous procedure to define the LPT. Although an embryonal version of the lattice momentum balance was already used in~\cite{ShanChen94}, the possibility of rigorously computing LPT from the momentum balance went unnoticed in all subsequent works. In this note, we delve deeper into the issue by showing how the momentum balance condition can apply in the SC-LBM framework, featuring different choices of pseudo-potentials and different forcing schemes. \\

The paper is organized as follows: in Sec.~\ref{sec:lbm}, essential features of the SC-LBM are recalled; in Sec.~\ref{sec:latticeMB}, we report on the analytical derivation of the lattice momentum balance with applications to different forcing schemes; numerical tests are discussed in Sec.~\ref{sec:res} and conclusions will follow in Sec.~\ref{sec:conclusions}.

%%%%%%%%%%%%%%%%%%%%%%%%%%%%%%%%%%%%%%%%%%%%%%%%%
\section{Shan-Chen Lattice Boltzmann Method}\label{sec:lbm}
%%%%%%%%%%%%%%%%%%%%%%%%%%%%%%%%%%%%%%%%%%%%%%%%%
LBMs consider the dynamics of discrete probability distribution functions (or populations), $f_{i}\left(\mathbf{x},t\right)$, representing the probability of finding a particle with discrete velocity $\boldsymbol{\xi}_{i}$ at the lattice location $\mathbf{x}$ at time $t$. The LBM evolution dynamics of $f_{i}\left(\mathbf{x},t\right)$ on a unitary time-lapse ($\Delta t=1$) reads as
\begin{equation}\label{eq:lbm}
\begin{split}
    f_{i}\left(\mathbf{x}+\boldsymbol{\xi}_{i},t+1\right) &- f_{i}\left(\mathbf{x},t\right) =\\
    & -\frac{1}{\tau}\left[f_{i}\left(\mathbf{x},t\right)- f_{i}^{(\text{eq})}\left(\mathbf{x},t\right)\right] + \Delta f_i(\textbf{x}, t),
\end{split}
\end{equation}
where the l.h.s. represents the populations streaming while the r.h.s. is composed of the Bhatnagar-Gross-Krook (BGK)~\cite{Bhatnagar_1954} collision operator, representing the relaxation of the populations $f_i$ towards the equilibrium populations $f_{i}^{(\text{eq})}$ with characteristic time $\tau$. The collision step is supplemented with a source term $\Delta f_i(\textbf{x}, t)$. The equilibrium populations depend on space and time via the local density $n\left(\mathbf{x},t\right)=\sum_i f_i\left(\mathbf{x},t\right)$ and the equilibrium velocity ${\bm u}^{(\text{eq})}\left(\mathbf{x},t\right)$ and they are obtained as a second-order approximation of the Maxwell distribution (repeated indices imply summation)
\begin{equation}\label{eq:feq}
\small 
  f_{i}^{(\text{eq})}=w_{i}n\left(1+\frac{\xi_{i}^{\alpha}u_{\alpha}^{(\text{eq})}}{c_{s}^{2}}+\frac{(\xi_{i}^{\alpha}\xi_i^{\beta}-\delta_{\alpha \beta} c_s^2) u_{\alpha}^{(\text{eq})} u_{\beta}^{(\text{eq})}}{2 c_s^4} 
  \right),
\end{equation}
where $w_i$ represent suitable weights with the property $\sum_i w_i=1$, and $c_s$ is isothermal speed of sound in lattice units with the property $\sum_i w_i \xi_{i}^{\alpha}\xi_{i}^{\beta}=c_s^2 \delta_{\alpha \beta}$, with $\delta_{\alpha \beta}$ the Kroneker delta.\\

Non-ideal effects are introduced via a non-ideal SC force, i.e., a force computed on lattice nodes that reads 
\begin{equation}\label{eq:SCForce}
    {\bm F}(\mathbf{x},t) = -Gc_s^2\,\psi(\mathbf{x},t)\sum_{\ell} W(|\boldsymbol{e}_i|^2)\, \psi(\mathbf{x} + \boldsymbol{e}_{\ell},t)\, \boldsymbol{e}_{\ell},
\end{equation}
where $\psi(\textbf{x}, t) = \psi(n(\textbf{x}, t))$ is the pseudo-potential function, i.e., a local function of the density $n$, implicitly depending on space and time. The coupling constant $G$ sets the strength of non-ideal interactions. The SC force couples the pseudo-potential function in the location $\textbf{x}$ with that of neighboring nodes identified via a suitable set of discrete directions $\boldsymbol{e}_{\ell}$. Such lattice coupling is weighted with statistical weights $W(|\boldsymbol{e}_{\ell}|^2)$ ensuring the desired level of isotropy conditions~\cite{Shan06,Sbragaglia07}. The SC force given in Eq.~\eqref{eq:SCForce} enters the LBM dynamics [see Eq.~\eqref{eq:lbm}] via the source term $\Delta f_i$ and the equilibrium velocity ${\bm u}^{(\text{eq})}$, depending on the forcing scheme used in the LBM. In this paper, we will consider three different forcing schemes: the Guo scheme~\cite{Guo2002}, the Shan-Chen (SC) scheme~\cite{ShanChen93, ShanChen94} and the Kupershtokh (Kup) scheme~\cite{kupershtokh2004new,kupershtokh2010criterion,kupershtokh2018lattice}. Details on these forcing schemes will be given later. Importantly, when a non-ideal force is added to the LBM dynamics, the hydrodynamical momentum (hereafter indicated with $n {\bm u}$) becomes a function of the forcing itself as
\begin{equation}
n(\textbf{x},t) {\bm u} (\textbf{x},t) = \sum_{i} f_i (\textbf{x},t) \boldsymbol{\xi}_i+\frac{{\bm F}}{2}.
\end{equation}
%%%%%%%%%%%%%%%%%%%%%%%%%%%%%%%%%%%%%%%%%%%%%%%%%%%%
\section{Lattice Momentum Balance}\label{sec:latticeMB}
\begin{figure*}[t!]
  \centering
    %\begin{tabular}{c c}
       \includegraphics[width=.96\linewidth]{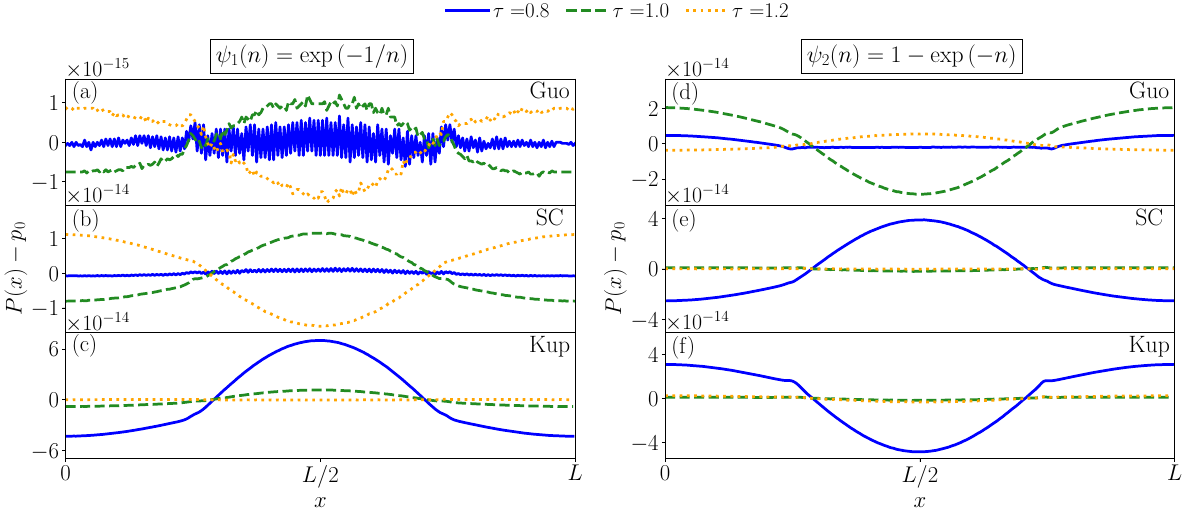}   %& \includegraphics[width=.48\linewidth]{figures/LPT_psi2.pdf} \\
    %\end{tabular}
  \caption{We report the deviations of the LPT, $P(x)$, with respect to the bulk pressure, $p_0$, along a flat interface developing along the $x$-coordinate, within the system domain $[0,L]$. In panels (a)-(c), we show data for simulations obtained with the pseudo-potential $\psi_1(n) = \exp{(-1/n)}$, while in panels (d)-(f) we show data for simulations obtained with the pseudo-potential $\psi_2(n) = 1-\exp{(-n)}$. Different forcing schemes are reported: the Guo forcing scheme (panels (a),(d)); the Shan-Chen (SC) forcing scheme (panels (b),(e)); the Kupershtokh (Kup) forcing scheme (panels (c),(f)). Different colors/line-styles correspond to different values of the relaxation time $\tau$. Data refer to the case with $G_c/G \simeq 0.94$, with $G$ being the coupling strength and $G_c$ its critical value (see Table~\ref{table}).
  \label{fig:LPTconstant}}
\end{figure*}

%%%%%%%%%%%%%%%%%%%%%%%%%%%%%%%%%%%%%%%%%%%%%%%%%%%%
By using a one-dimensional formulation, let us derive the lattice momentum balance equation~\cite{Guo_2011, Guo_2021}. Let us start from the lattice Boltzmann equation at steady state
\begin{equation}\label{eq:LBMD1Q3}
    f_{i}\left(x+\xi_{i}\right)-f_{i}\left(x\right)=-\frac{1}{\tau}\left[f_{i}\left(x\right)-f_{i}^{\text{(eq)}}\left(x\right)\right]+\Delta f_{i}\left(x\right).
\end{equation}
We consider the D1Q3 LBM scheme with discrete velocities $\left\{ \xi_{0}=0,\xi_{1}=1,\xi_{2}=-1\right\}$ with weights $\left\{ w_{0}=2/3,w_{1}=w_{2}=1/6\right\}$. Out of the normalization condition, $w_0+2w_1=1$, and the second order moment, $\sum_i w_i \xi_i^2 = c_s^2=1/3$, we find the relation between  the weights and $c_s^2$, i.e., $w_1=c_s^2/2$ and 
\begin{equation}\label{eq:weight}
w_0=1-c_s^2.
\end{equation}
Notice that the D1Q3 is also an \emph{optimal} Gaussian quadrature~\cite{shan2016the}. In general, we can write the weights as a function of the speed of sound, keeping in mind that the stencil is not necessarily optimal, as the latter feature depends on the value of $c_s^2$, i.e., only for $c_s^2=1/3$ for D1Q3.\\ 
In all subsequent developments, we also assume to deal with a single-range potential, i.e., we use the SC form given in Eq.~\eqref{eq:SCForce} where the summation considers the D1Q3 discrete directions used in Eq.~\eqref{eq:LBMD1Q3}. Then, the forcing reads
\begin{equation}\label{eq:SCforce1d}
F\left(x\right)=-\frac{Gc_{s}^{2}}{2}\psi\left(x\right)\left[\psi\left(x+1\right)-\psi\left(x-1\right)\right] \ ,
\end{equation}
where we used the projection of the single-belt forcing stencil $W(1) = W(2) = 1/2$~\cite{Lulli_2021}. To progress, we need to determine the populations $f_0$, $f_1$, and $f_2$. Let us start with $f_{0}$, associated with $\xi_{0}=0$, which is non-streaming on the lattice: it can be determined by the associated component of the steady-state equation as (space dependency is not reported to keep compact notation): 
\begin{equation}\label{eq:resultf0}
f_{0}=f_{0}^{\text{(eq)}}+\tau \Delta f_{0}=w_0 n-n (u^{\text{(eq)}})^2+\tau \Delta f_{0} \ ,
\end{equation}
where we used that $f^{(\text{eq})}_0=w_0 n-w_0 n (u^{\text{(eq)}})^2/2c_s^2=w_0 n-n (u^{\text{(eq)}})^2$. In order to determine the other two populations, namely $f_{1}$ and $f_{2}$, we resort to the local definition of density $n$ and hydrodynamical momentum $nu$, respectively, so that
\begin{equation}
    \begin{cases}
\begin{split} & n-f_{0}=f_{1}+f_{2}\\
 & n u-\frac{F}{2}=f_{1}-f_{2}
\end{split}
\end{cases}
\end{equation}
yielding as solution
\begin{equation*}
\begin{cases}
\begin{split} & f_{1}=\frac{1}{2}\left[n-f_{0}+n u-\frac{F}{2} \right]\\
 & f_{2} =\frac{1}{2}\left[n-f_{0}-n u+\frac{F}{2} \right]
\end{split}
\end{cases}
\end{equation*}
which after substitution of Eq.~\eqref{eq:resultf0} reads
\begin{equation}\label{eq:f1f2}
    \begin{cases}
\begin{split} & f_{1}=\frac{1}{2}\left[c_s^2 n+nu+n (u^{(\text{eq})})^2-\tau \Delta f_0 -\frac{F}{2}\right]\\
 & f_{2}=\frac{1}{2}\left[c_s^2 n-n u+n (u^{(\text{eq})})^2-\tau \Delta f_0 +\frac{F}{2}\right]
\end{split}
\end{cases}
\end{equation}
where we have used Eq.~\eqref{eq:weight}. We can go back to the steady-state equation [see Eq.~\eqref{eq:LBMD1Q3}] and compute its first moment. At this stage, we require the constraint that the first order moment of the r.h.s. of Eq.~\eqref{eq:LBMD1Q3} equals the forcing $F$~\cite{Guo2002}, i.e.,
\begin{equation}
\sum_{i=0}^{2}\left(-\frac{1}{\tau}\left[f_{i}-f_{i}^{\text{(eq)}}\right]+\Delta f_{i}\right)\xi_i=F ,\  
\end{equation}
and hence, we get (explicitly reporting the space dependence again)
\begin{equation}\label{eq:final}
f_{1}\left(x+1\right)-f_{2}\left(x-1\right)-n(x) u(x)-\frac{F(x)}{2}=0.
\end{equation}
Then, we write the equilibrium velocity as the sum of the hydrodynamical velocity $u(x)$ and a correction term $\Delta u(x)$
\begin{equation}
u^{(\text{eq})}(x)=u(x)+\Delta u(x)
\end{equation}
and, under the assumption of equilibrium with a zero hydrodynamical velocity~\cite{Guo_2021}, $u=0$, Eqs.~\eqref{eq:final} and~\eqref{eq:f1f2} yield
\begin{equation}\label{eq:LMBfinal}
\begin{split} & \frac{c_{s}^{2}}{2}\left[n\left(x+1\right)-n\left(x-1\right)\right]\\
- & \frac{1}{4}\left[F\left(x+1\right)+2F\left(x\right)+F\left(x-1\right)\right]\\
- & \frac{\tau}{2} \left[ \Delta f^{(u=0)}_{0}(x+1) - \Delta f^{(u=0)}_0(x-1)\right]\\
+ & \frac{1}{2} \left[ n(x+1)(\Delta u(x+1))^2-n(x-1)(\Delta u(x-1))^2 \right]=0.
\end{split}
\end{equation}
The first term in square brackets of Eq.~\eqref{eq:LMBfinal} represents the finite difference approximation of the derivative of the bulk pressure of an ideal gas $\frac{c_{s}^{2}}{2}\left[n\left(x+1\right)-n\left(x-1\right)\right]\simeq\partial_{x}\left(c_{s}^{2} n\right)$, while the second term in square brackets can be seen as an average force over neighboring points. Interestingly, in Ref.~\cite{Shan08} Shan proposed to compute the force crossing a given area element exactly using an average force defined according to some specific geometric rules~\cite{Shan08,SbragagliaBelardinelli13,Lulli_2021}. The last two terms in Eq.~\eqref{eq:LMBfinal} are contributions specific to the forcing implementation.\\
Based on the equation for the SC force in Eq.~\eqref{eq:SCforce1d}, the second term in Eq.~\eqref{eq:LMBfinal} reads
\begin{equation}\label{eq:unrolling_F}
    \begin{split}  - &\frac{1}{4}  \left[F\left(x+1\right)+2F\left(x\right)+F\left(x-1\right)\right]=\\
  +&\frac{Gc_{s}^{2}}{8}\psi\left(x+1\right)\left[\psi\left(x+2\right)+\psi\left(x\right)\right]\\
- & \frac{Gc_{s}^{2}}{8}\psi\left(x-1\right)\left[\psi\left(x\right)+\psi\left(x-2\right)\right],
\end{split}
\end{equation}
so that Eq.~\eqref{eq:LMBfinal} can be written as an exact finite difference equation for an LPT 
\begin{equation}\label{eq:prefactor12}
\frac{1}{2}\left[P(x+1)-P(x-1)\right]=0
\end{equation}
where 
\begin{equation}\label{eq:LPTgeneral}
\begin{split}
P(x) = & c_s^2 n(x)+\frac{Gc_{s}^{2}}{4}\psi\left(x\right)\left[\psi\left(x+1\right)+\psi\left(x-1\right)\right]\\
+ & n(x) (\Delta u(x))^2-\tau \Delta f^{(u=0)}_0(x)
\end{split}
\end{equation}
is the normal component of the interaction pressure tensor~\cite{Shan08,SbragagliaBelardinelli13, Lulli_2021}. This result provides an obvious explanation as to why the normal component of the SC-LPT implements the mechanical equilibrium condition~\cite{SbragagliaBelardinelli13, Lulli_2021} to machine precision across a flat interface on the lattice [see Fig.~\ref{fig:LPTconstant}]: this happens because the mechanical equilibrium condition is equivalent to the first moment of the lattice Boltzmann equation, which is exactly implemented in the simulations. The two terms $n (\Delta u)^2$ and $-\tau \Delta f^{(u=0)}_0$ in Eq.~\eqref{eq:LPTgeneral} depend on the specific forcing implementation. In the following section, we report the LPT analysis for three different forcing implementation schemes. We remark that for the cases here analyzed $\Delta f^{(u=0)}_0$ always vanishes: this seems reasonable because ({\it i}) due to mass conservation, the zeroth moment of $\Delta f_i$ must vanish and ({\it ii}) the second moment of $\Delta f_i$ needs to be proportional to $u F$ to remove lattice effects in the hydrodynamic limit~\cite{Guo2002, Wagner_2006}. Furthermore, one might wish to adopt the present analysis for settings where no such constraints are present , e.g., when modeling diffusive dynamics ~\cite{Biferale2011}, so that formally keeping $\Delta f^{(u=0)}_0$ renders the results more general.
%%%%%%%%%%%%%%%%%%%%%%%%%%%%%%%
\subsection{Guo forcing scheme}\label{subsec:GUO}
%%%%%%%%%%%%%%%%%%%%%%%%%%%%%%%%
In the Guo forcing scheme the equilibrium velocity $u^{(\text{eq})}$ and the source term $\Delta f_i$ are evaluated as~\cite{Guo2002}:
$$
u^{(\text{eq})}=u
$$
$$
\Delta f_{i} =\left(1-\frac{1}{2\tau}\right)w_{i} \left[\frac{\xi_i}{c_{s}^{2}} + \frac{\xi_i^2-c_s^2}{c_{s}^{4}} u\right]F
$$
so that $\Delta f_0 = (c_s^2-1)(1-\frac{1}{2\tau})uF/c_s^2$. This
means that in equilibrium ($u=0$), we have
\begin{equation}
\Delta u=\Delta f^{(u=0)}_0=0
\end{equation}
and hence Eq.~\eqref{eq:LPTgeneral} results in
\begin{equation}\label{eq:LPT_GUO}
\begin{split}
P(x) = & c_s^2 n(x)+\frac{Gc_{s}^{2}}{4}\psi\left(x\right)\left[\psi\left(x+1\right)+\psi\left(x-1\right)\right],
\end{split}
\end{equation}
meaning that in the Guo forcing scheme, the LPT recovered from the momentum balance equations coincides with the geometric prediction proposed by Shan~\cite{Shan08}. 
%%%%%%%%%%%%%%%%%%%%%%%%%%%%%%%%%%%%%%%%%%%%%%%%%%%%%%%%%%%%%%%%%%%%%%%%%%%%%%%%%%%%%%%%%%%%%%%%%%
%%%%%%%%%%%%%%%%%%%%%%%%%%%%%%%%%%%%%%%%%%%%%%%%%%%%%%%%%%%%%%%%%%%%%%%%%%%%%%%%%%%%%%%%%%%%%%%%%%
%%%%%%%%%%%%%%%%%%%%%%%%%%%%%%%%%%%%%%%%%%
\subsection{Shan-Chen (SC) forcing scheme}\label{subsec:SC}
%%%%%%%%%%%%%%%%%%%%%%%%%%%%%%%%%%%%%%%%%%
In the SC forcing scheme, $\Delta f_i = 0$ by construction and the equilibrium velocity appearing in Eq.~\eqref{eq:feq} is computed from the pre-collisional momentum $nu^{\text{(pre)}}$ shifted with a term which is proportional to the forcing times the relaxation time~\cite{ShanChen93,ShanChen94}. In terms of the hydrodynamical momentum $n u$, the equilibrium velocity is expressed in compact notation as:
$$ 
nu^{\text{(eq)}}=nu^{\text{(pre)}}+\tau F = n u+\left(\tau-\frac{1}{2}\right) F.
$$
Furthermore, the source term $\Delta f_i$ is null in the SC forcing scheme. This leads to
\begin{equation}
\Delta u=\left(\tau-\frac{1}{2}\right) \frac{F}{n}  \ , \hspace{.2in} \Delta f^{(u=0)}_0=0
\end{equation}
hence, Eq.~\eqref{eq:LPTgeneral} results in
\begin{equation}\label{eq:LPT_SC}
\begin{split}
P\left(x\right)  = c_s^2 n(x) & + \frac{Gc_{s}^{2}}{4}\psi\left(x\right)\left[\psi\left(x+1\right)+\psi\left(x-1\right)\right] \\ & +\left(\tau-\frac{1}{2}\right)^{2}\frac{F^2(x)}{n(x)}.
\end{split}
\end{equation}
It is interesting to note that in Ref.~\cite{ShanChen94}, the pressure tensor was already analyzed by leveraging the stationarity condition, allowing us to write the equilibrium velocity exactly in terms of lattice quantities [see Eq.(16) in Ref.~\cite{ShanChen94}]. This expression was then considered in the small-gradient approximation yielding the Taylor expansion of the entire pressure tensor [see Eq.(19) in Ref.~\cite{ShanChen94}], whose one-dimensional projection would exactly match the expansion of Eq.~\eqref{eq:LPT_SC} of the present work, i.e., the $F^2$ contribution was already captured in~\cite{ShanChen94}. Indeed, Taylor expansion was necessary to obtain the entire pressure tensor rather than only the normal component. 

Notice that the LPT obtained with the SC forcing scheme is the same as the LPT obtained with the Guo forcing scheme in Eq.~\eqref{eq:LPT_GUO}, except for an additional contribution which disappears when $\tau\approx 1/2$ and/or for low amplitudes of the force ($F \ll 1$). Yet, in the context of multi-phase flow simulations, it is very likely that these deviation terms will be non-negligible since the Reynolds number is generally relatively small and the force amplitude is quite large at the interface.
%%%%%%%%%%%%%%%%%%%%%%%%%%%%%%%%%%%%%%%%%%%%%
\subsection{Kupershtokh (Kup) forcing scheme}\label{subsec:Kup}
%%%%%%%%%%%%%%%%%%%%%%%%%%%%%%%%%%%%%%%%%%%%%
In the Kupershtokh (Kup) forcing scheme, body forces are introduced via the exact difference method~\cite{kupershtokh2004new,kupershtokh2010criterion,kupershtokh2018lattice}. The equilibrium velocity is set to the pre-collisional velocity ($u^{\text{pre}}$), hence the equilibrium velocity can be expressed in compact form in terms of the hydrodynamical momentum $n u$ as:
\begin{equation*}
    nu^{\text{(eq)}}=nu^{\text{(pre)}}= n u-\frac{F}{2}.
\end{equation*}
For the source term, we have
\begin{equation*}
    \Delta f_{i}=f_{i}^{\text{(eq)}}\left(n,u^{\text{(eq)}}+\frac{F}{n}\right)-f_{i}^{\text{(eq)}}\left(n,u^{\text{(eq)}}\right)
\end{equation*}
and, given the expression of $f^{(\text{eq})}_0$ given in Eq.~\eqref{eq:resultf0}, we get
$$
\Delta f_0 = n \left(u-\frac{F}{2n} \right)^2 - n \left(u+\frac{F}{2n} \right)^2 = uF.
$$
Hence, in equilibrium ($u=0$), we have
\begin{equation}
\Delta u=-\frac{F}{2n} \ , \hspace{.2in} \Delta f^{(u=0)}_0=0
\end{equation}
and hence Eq.~\eqref{eq:LPTgeneral} results in
\begin{equation}\label{eq:LPT_KUP}
\begin{split}
P\left(x\right)= c_s^2 n(x) & +\frac{Gc_{s}^{2}}{4}\psi\left(x\right)\left[\psi\left(x+1\right)+\psi\left(x-1\right)\right]\\
& + \frac{1}{4}\frac{F^2(x)}{n(x)} \ ,
\end{split}
\end{equation}
which coincides with the LPT obtained from the SC forcing scheme [see Eq.~\eqref{eq:LPT_SC}] when $\tau=1$.
%%%%%%%%%%%%%%%%%%%%%%%%%%%%%%%%%%%%%%%%
\section{Numerical Tests}\label{sec:res}
%%%%%%%%%%%%%%%%%%%%%%%%%%%%%%%%%%%%%%%%
%%%%%%%%%%%%%%%%%%%%%%%%%%%%%%%%%%%%%%%%%%%%%%%%%%%%%%%%%%%%%%%%%%%%%%%%%%%%%%%%%%%%%%%%%%%%%%%%%%
\begin{figure*}[th!]
\centering
\includegraphics[width=.9\linewidth]{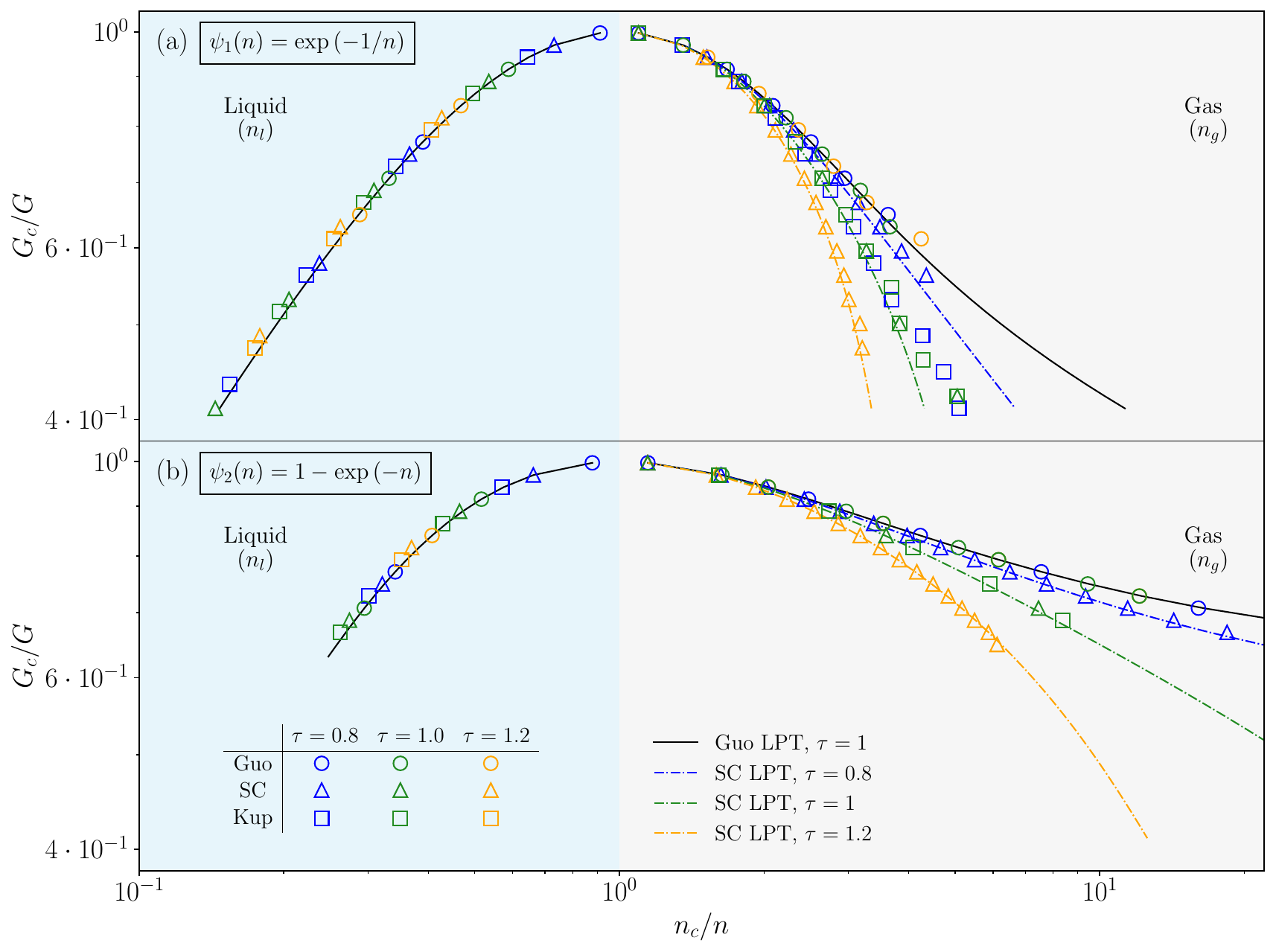}
\caption{Coexistence curves of the inverse rescaled density $n_c/n$ as a function of the inverse rescaled coupling constant $G_c/G$. The quantities $n_c$ and $G_c$ represent the critical density and critical coupling constant, respectively. Empty points represent simulation data, with different forcing implementations indicated with different symbols, while data for different values of $\tau$ are indicated with different colors. The LPT predictions for the SC forcing scheme are shown as dashed-dotted lines. The black solid line refers to the LPT prediction for the Guo scheme. In panel (a), we show simulation data and LPT predictions obtained with the pseudo-potential $\psi_1(n) = \exp{(-1/n)}$, while in panel (b) we report on the case with $\psi_2(n) = 1-\exp{(-n)}$.}\label{fig:coex}
\end{figure*}
%%%%%%%%%%%%%%%%%%%%%%%%%%%%%%%%%%%%%%%%%%%%%%%%%%%%%%%%%%%%%%%%%%%%%%%%%%%%%%%%%%%%%%%%%%%%%%%%%%
To confirm the validity of the given definition of the LPT, we performed two-dimensional simulations of a flat interface developing along the $x$-coordinate in a system domain with extension $L$. Numerical simulations have been conducted for the three different forcing implementations analyzed in the previous section and for two different definitions of the pseudo-potential: (\textit{i}) $\psi_1(n) = \exp{(-1/n)}$ and (\textit{ii}) $\psi_2(n) = 1-\exp{(-n)}$. In numerical simulations, we established that the system reaches its equilibrium state when the average of the hydrodynamic velocity assumes values smaller than $10^{-13}$. We first verified that the LPTs $P(x)$ given in Eqs.~\eqref{eq:LPT_GUO},~\eqref{eq:LPT_SC} and~\eqref{eq:LPT_KUP} are constant along the interface. Results are reported in Fig.~\ref{fig:LPTconstant}, where we show the difference between $P(x)$ and the bulk pressure $p_0$,  displaying fluctuations in the order of ${\cal O}(10^{-15}-10^{-14})$. Indeed, one can observe that fluctuations in Fig.~\ref{fig:LPTconstant} display a spurious $\tau$ dependence for both Guo and Kupershtokh forcing schemes. These discrepancies are likely caused by the fact that the lattice momentum balance is valid at steady state, i.e., dynamically, with the repeated application of the lattice Boltzmann equation on the lattice populations, so that cancellations are affected by numerical rounding carrying a dependence on the actual value of the parameters, including $\tau$. We then measured the equilibrium bulk densities for the gas and liquid phases ($n_g$ and $n_l$) by changing the coupling constant $G$ in Eq.~\eqref{eq:SCForce}. Numerical results are compared with theoretical predictions for $n_g$ and $n_l$ obtained by following quite standard procedures~\cite{Shan08,Lulli_2021}. In a nutshell, starting from the mechanical equilibrium condition $P(x)=p_0$, we perform a Taylor expansion of the LPT to get the following type of differential equation  
\begin{equation}\label{eq:embrionalMaxwell}
p_0-c_s^2 n-\frac{G c_s^2}{2}\psi^2=\beta \psi \frac{d^2 \psi}{d x^2}+\alpha \left(\frac{d \psi}{dx}\right)^2
\end{equation}
where the coefficients $\alpha$ and $\beta$ depend on the coupling constant $G$ and the forcing scheme used, hence also on $\tau$ when applicable. It is then possible to relate the squared derivative of the density $n$ to a suitable integral evaluated between $n_g$ and $n$ \footnote{This is based on the identities $\frac{d^2 \psi}{dx^2}=\frac{1}{2}\frac{d}{d \psi} \left(\frac{d\psi}{dx}\right)^2$ and $\frac{\beta}{2}\psi^{\epsilon+1}\frac{d}{d\psi}\left[\frac{1}{\psi^{\epsilon}}\left(\frac{d \psi}{dx} \right)^2 \right]  = \frac{\beta}{2}\psi \frac{d}{d \psi} \left(\frac{d \psi}{dx} \right)^2+\alpha \left(\frac{d \psi}{dx} \right)^2$} 
$$
\left(\frac{dn}{dx} \right)^2= \frac{2 \psi^{\epsilon}}{\beta (\psi^{\prime})^2} \int_{n_g}^{n} d\bar{n} \frac{\psi^{\prime}}{\psi^{\epsilon+1}} \left(p_0-c_s^2 \bar{n}-\frac{G c_s^2}{2}\psi^2(\bar{n}) \right)
$$
with $\epsilon=-2 \alpha/\beta$. Hence, from the condition of null derivative in the bulk phases, we come up with an integral constraint to determine both $n_g$ and $n_l$
\begin{equation}\label{eq:Maxwell}
\int_{n_g}^{n_l} d\bar{n} \frac{\psi^{\prime}}{\psi^{\epsilon+1}} \left[p_0-c_s^2 \bar{n}-\frac{G c_s^2}{2}\psi^2(\bar{n}) \right]=0
\end{equation}
that is used to evaluate both $n_g$ and $n_l$ once the coupling constant $G$ is set.  The approaching numerical instability for large values of the system domain and $G$ has led us to choose the system size by rescaling it with the distance from the critical point. 

In Fig.~\ref{fig:coex}, we report the coexistence curves for different values of $G$. The bulk densities $n_{g,l}$ and the coupling constant $G$ are inversely rescaled with the critical density $n_c$ and the critical strength $G_c$, respectively. Note that the values of $n_c$ and $G_c$, as well as the critical pressure ($p_c$), are different between the two pseudo-potentials [see Table~\ref{table}].
%%%%%%%%%%%%%%%%%%%%%%%%%%%%%%%%%%%%%%%%%%%%%%%%%%%%%%%%%%%%%%%%%%%%%%%
\begin{table}[t!]
\centering
%    \begin{tabular}{|m{1.1cm}| m{3.2cm} m{3.2cm}|}
    \begin{tabular}{|c| c | c |}
    \hline 
 & \multicolumn{2}{c|}{Pseudo-potential} \\
\hline
     &\hspace{0.2cm} $\psi_1(n) = \exp{(-1/n)}$ \hspace{0.2cm} & \hspace{0.1cm} $\psi_2(n)= 1-\exp{(-n)}$ \hspace{0.1cm} \\ 
    \hline
    \hspace{0.2cm} $n_c$ \hspace{0.2cm} & $1$ & $\log(2) \simeq 0.693$\\
%\hline
    $G_c$ & $-e^2\simeq-2.46$ & $-4$\\
%    \hline
    $p_c$ & $c_s^2/2\,\simeq 0.167$ & $c_s^2(\log(2) - 1/2) \simeq 0.0644$\\
 \hline
    \end{tabular}
    \caption{Values for $n_c$, $G_c$ and $p_c$ for the two definitions of the pseudo-potential $\psi$ under investigation.}
    \label{table}
\end{table}
%%%%%%%%%%%%%%%%%%%%%%%%%%%%%%%%%%%%%%%%%%%%%%%%%%%%%%%%%%%%%%%%%%
In panel (a), we show results obtained with the pseudo-potential $\psi_1$, while in panel (b), we show results obtained with the pseudo-potential $\psi_2$. Although we systematically varied the relaxation time value $\tau$ in the range [0.8-1.2], for the sake of clarity, we report data for three representative values only, i.e., $\tau = 0.8, \ 1.0, \ 1.2$. Simulation data (empty points) are shown with different symbols (colors) corresponding to different forcing schemes (values of $\tau$). Dashed-dotted lines draw the LPT predictions for the SC forcing scheme [see Eq.~\eqref{eq:LPT_SC}]. The continuous black line indicates the LPT prediction for the Guo forcing scheme [see Eq.~\eqref{eq:LPT_GUO}]. Fig.~\ref{fig:coex} confirms that for all values of $\tau$, data for the Kupershtokh forcing scheme falls on the SC LPT curve for $\tau =1.0$. Furthermore, coexistence curves exhibit a good agreement between simulation data and LPT predictions for both definitions of $\psi$. The latter means that the control of the system's thermodynamic behavior starting from the LPT is robust. Discrepancies observed at a large value of $G$ are probably due to the fact that the theoretical analysis leading to Eq.~\eqref{eq:Maxwell} is based on a truncated Taylor expansion of the LPT; hence, we expect higher order in the expansion to matter when density gradients increase at the interface, i.e., when $G$ increases.\\
%%%%%%%%%%%%%%%%%%%%%%%%%%%%%%%%%%%%%%%%%%%%
\section{Conclusions}\label{sec:conclusions}
%%%%%%%%%%%%%%%%%%%%%%%%%%%%%%%%%%%%%%%%%%%%
We analyzed the lattice momentum balance for the Shan-Chen (SC) multi-phase lattice Boltzmann model (LBM). For a simple 1D case with single range potential, we showed that this analysis allows to rigorously construct \emph{a priori} the model's lattice pressure tensor (LPT) without arbitrariness and consistently with the LBM, nor resorting to a geometric construction~\cite{Shan08,Lulli_2021}. Interestingly, the lattice momentum balance approach is quite robust as it can also be generalized to different forcing implementation schemes~\cite{Guo2002,ShanChen94,Kupershtokh_2009}.

Given that the range of the forcing stencil is independent of that of the discrete velocity stencil, it is possible to successfully use Eq.~\eqref{eq:LMBfinal} also in the case of multi-range interactions. One could also consider extending the method to a higher-dimensional case: a two-dimensional extension, for example, would amount to considering a \emph{tilted} interface with respect to the underlying lattice axes by an angle $\theta$. However, in this case, one could define velocity vectors parallel to the interface, only in the cases of $\theta=0$, $\theta=\pi/2$, and $\theta=\pi/4$. While the first two cases correspond to the present work, the latter has not yet been explored. For any other value of $\theta$, one would need to resort to some approximation, e.g., to Chapman-Enskog~\cite{kruger2017lattice}.\\ 
The effects of higher-order terms, in the form of higher-order moments, can also be investigated for both the equilibrium~\cite{COREIXAS_PRE_100_2019} and the forcing approach~\cite{DEROSIS_PoF_31_2019}. However, since the lattice momentum balance calculations are based on the computation of zeroth and first moments, with the interface assumed to be steady and at rest ($u=0$), it is not surprising that the final result (not reported here) remains independent of any higher-order contributions. Similar independence with respect to the potential range was already pointed out in Ref.~\cite{lulli2024metastable}, where the lattice pressure tensor allowed for high-precision modeling of the hydrodynamic fluctuations in the multi-phase SC model for single- and multi-range potentials.\\ 
Finally, it is also interesting to notice that the lattice momentum balance allows us to define either a \emph{lattice chemical potential}, as it is the case in Ref.~\cite{Guo_2021}, or a lattice pressure tensor, as it is shown in the present work. Following the arguments proposed in Ref.~\cite{Wagner_2006}, one may try to follow the path of exact lattice relations in order to define a model where the lattice chemical potential and the lattice pressure tensor are connected by some discrete version of the divergence operator, which would eventually lead to a fully thermodynamic consistent approach.

\begin{acknowledgements}
We sincerely thank Giacomo Falcucci for his valuable discussions and insightful suggestions. FP and MS acknowledge the support of the National Center for HPC, Big Data and Quantum Computing, Project CN\_00000013 – CUP E83C22003230001 and CUP B93C22000620006, Mission 4 Component 2 Investment 1.4, funded by the European Union – NextGenerationEU.  
\end{acknowledgements}

The data that support the findings of this study are available from the corresponding author upon reasonable request.

\bibliographystyle{apsrev4-2}
\bibliography{biblio}
\end{document}